\documentclass[11pt]{article}  
\newcommand{\ieee}[1]{1}\newcommand{\nonieee}[1]{#1}   

\nonieee{\newenvironment{proof}{\noindent {\em Proof}.\ }{\hspace*{\fill}$\halmos$\medskip}}

\nonieee{\newenvironment{keywords}{\noindent {\em Keywords:}.\ }{\hspace*{\fill}\medskip}}

\usepackage{amsfonts,amssymb,curves}

\usepackage{makeidx}     
\usepackage{graphicx}    

\newcommand{\R}{{\mathbb R}}  
\newcommand{\C}{{\mathbb C}}  

\newtheorem{theorem}{Theorem}
\newtheorem{itlemma}{Lemma}[section] 
\newtheorem{itproposition}[itlemma]{Proposition}
\newtheorem{itcorollary}[itlemma]{Corollary}
\newtheorem{itremark}[itlemma]{Remark}
\newtheorem{itdefinition}[itlemma]{Definition}
\newtheorem{itexample}[itlemma]{Example}

\newenvironment{lemma}{\begin{itlemma}\rm}{\end{itlemma}} 
\newenvironment{remark}{\begin{itremark}\rm}{\end{itremark}} 
\newenvironment{corollary}{\begin{itcorollary}\rm}{\end{itcorollary}}
\newenvironment{proposition}{\begin{itproposition}\rm}{\end{itproposition}}
\newenvironment{definition}{\begin{itdefinition}\rm}{\end{itdefinition}}
\newenvironment{example}{\begin{itexample}\rm}{\end{itexample}}

\newcommand{\text}[1]{\hbox{\rm \ #1\ \/}}
\newcommand{\be}[1]{\begin{equation}\label{#1}}
\newcommand{\ee}{\end{equation}}
\newcommand{\beqn}{\begin{eqnarray*}}
\newcommand{\eeqn}{\end{eqnarray*}}
\newcommand{\beq}{\begin{eqnarray}}
\newcommand{\eeq}{\end{eqnarray}}
\newcommand{\bl}[1]{\begin{lemma}\label{#1}}
\newcommand{\ble}[1]{\begin{lemmaex}\label{#1}}
\newcommand{\br}[1]{\begin{remark}\label{#1}}
\newcommand{\bt}[1]{\begin{theorem}\label{#1}}
\newcommand{\bd}[1]{\begin{definition}\label{#1}}
\newcommand{\bp}[1]{\begin{proposition}\label{#1}}
\newcommand{\bc}[1]{\begin{corollary}\label{#1}}
\newcommand{\bfact}[1]{\begin{fact}\label{#1}}
\newcommand{\ber}[1]{\begin{exercise}\label{#1}}
\newcommand{\bex}[1]{\begin{example}\label{#1}}
\newcommand{\bem}[1]{\begin{example}\label{#1}}  
\newcommand{\ec}{\mybox\end{corollary}}
\newcommand{\efact}{\mybox\end{fact}}
\newcommand{\eer}{\mybox\end{exercise}}
\newcommand{\eex}{\mybox\end{example}}
\newcommand{\eem}{\mybox\end{example}}
\newcommand{\el}{\mybox\end{lemma}}
\newcommand{\ele}{\mybox\end{lemmaex}}
\newcommand{\er}{\mybox\end{remark}}
\newcommand{\et}{\qed\end{theorem}}
\newcommand{\ed}{\mybox\end{definition}}
\newcommand{\ep}{\mybox\end{proposition}}
\newcommand{\epr}{\end{proof}}
\newcommand{\bpr}{\begin{proof}}

\newcommand{\ecs}{\end{corollary}}
\newcommand{\eers}{\end{exercise}}
\newcommand{\eexs}{\end{example}}
\newcommand{\eems}{\end{example}}
\newcommand{\els}{\end{lemma}}
\newcommand{\eles}{\end{lemmaex}}
\newcommand{\ers}{\end{remark}}
\newcommand{\ets}{\end{theorem}}
\newcommand{\eds}{\end{definition}}
\newcommand{\eps}{\end{proposition}}
\newcommand{\halmos}{\rule{1ex}{1.4ex}}

\newcommand{\qed}{\hfill \halmos} 
\newcommand{\mybox}{\hfill $\Box$} 

\newcommand{\ben}{\begin{enumerate}}
\newcommand{\een}{\end{enumerate}}

\newcommand{\bi}{\begin{itemize}}
\newcommand{\ei}{\end{itemize}}

\newcommand{\barM}{{\bar M}}

\date{}

\begin{document}

\title{Oscillations in I/O Monotone Systems\\
 under Negative Feedback}

\author{David Angeli and Eduardo D. Sontag%
\thanks{D. Angeli is with the
Dip. Sistemi e Informatica,
University of Florence, 50139 Firenze, Italy
(angeli@dsi.unifi.it)}%
\thanks{E.D. Sontag is with the
Dept. of Mathematics,
Rutgers University, NJ, USA.
(sontag@math.rutgers.edu)}}%
    \maketitle

\markboth{Monotone Systems under Negative Feedback}{Monotone Systems under Negative Feedback}

\begin{abstract}
  Oscillatory behavior is a key property of many biological systems.  The
  Small-Gain Theorem (SGT) for input/output monotone systems provides a
  sufficient condition for global asymptotic stability of an equilibrium and
  hence its violation is a necessary condition for the existence of periodic
  solutions.  One advantage of the use of the monotone SGT technique is its
  robustness with respect to all perturbations that preserve monotonicity and
  stability properties of a very low-dimensional (in many interesting
  examples, just one-dimensional) model reduction.  This robustness makes the
  technique useful in the analysis of molecular biological models in which
  there is large uncertainty regarding the values of kinetic and other
  parameters.  However, verifying the conditions needed in order to apply the
  SGT is not always easy.  This paper provides an approach to the verification
  of the needed properties, and illustrates the approach through an
  application to a classical model of circadian oscillations, as a nontrivial
  ``case study,'' and also provides a theorem in the converse direction of
  predicting oscillations when the SGT conditions fail.

\end{abstract}


\section{Introduction}

Motivated by applications to cell signaling, our previous
paper~\cite{monotoneTAC} introduced the class of monotone input/output
systems, and provided a technique for the analysis of negative feedback loops
around such systems.  The main theorem gave a simple graphical test which may
be interpreted as a monotone small gain theorem (``SGT'' from now on) for
establishing the global asymptotic stability of a unique equilibrium,
a stability that persists even under arbitrary transmission delays in the
feedback loop.
Since that paper, various papers have followed-up on these ideas, see for
example~\cite{%
malisoff-leenheer,
hybridmonotonefeedback,
bastin,
falugi,
enciso-dcds,
predatorpreysgt05,
enciso_smith_sontagJDE06,
testosterone,
angelileenheersontagSCL04,
gedeon05,
ejc05es}.
The present paper, which was presented in preliminary form at the 2004 IEEE
Conference on Decision and Control, has two purposes.

The first purpose is to develop explicit conditions so as to make it easier to
apply the SGT theorem, for a class of systems of biological significance, a
subset of the class of tridiagonal systems with inputs and outputs.  
Tridiagonal systems (with no inputs and outputs) were introduced largely
for the study of gene networks and population models, and many results are
known for them, see for 
instance~\cite{%
Sm,
smith-tridiagonal}.
Deep achievements of the theory include the generalization of the
Poincar\'e-Bendixson Theorem, from planar systems to tridiagonal systems of
arbitrary dimension, due to Mallet-Paret and Smith~\cite{MPSmith} as well as a
later generalization to include delays due to Mallet-Paret and
Sell~\cite{malletparet}. 
For our class of systems, we provide in Theorem~\ref{main-P-theo} sufficient
conditions that guarantee the existence of characteristics (``nonlinear DC
gain''), which is one of the ingredients needed in the SGT Theorem
from~\cite{monotoneTAC}.

Negative feedback is often associated with oscillations, and in that context
one may alternatively view the failure of the SGT condition as providing a
necessary condition for a system to exhibit periodic behaviors, and this is
the way in which the SGT theorem has been often applied.

The conditions given in Theorem~\ref{main-P-theo} arose from our analysis of a
classical model of circadian oscillations.  
The molecular biology underlying the circadian rhythm in {\em Drosophila\/}
is currently the focus of a large amount of both experimental and theoretical
work. 
The most classical model is that of Goldbeter, who proposed a simple model for
circadian oscillations in \emph{Drosophila}, 
see~\cite{goldbeter95} and the book~\cite{Goldbeter}.
The key to the Goldbeter model is the auto-inhibition of the transcription of
the gene {\em per\/}.  This inhibition is through a loop that involves
translational and post-transcriptional modifications as well as nuclear
translocation. 
Although, by now, several more realistic models are available, in particular
incorporating other genes, see e.g.~\cite{leloup1,leloup2}, this simpler model
exhibits many realistic features, such as a close to 24-hour period, and
has been one of the main paradigms in the study of oscillations in gene
networks.
Thus, we use Goldbeter's original model as our ``case study'' to illustrate the
mathematical techniques.

The second purpose of this paper is to further explore the idea that,
conversely, failure of the SGT conditions may lead to oscillations if there is
a delay in the feedback loop.  (As with the Classical Small Gain Theorem, of
course the SGT is far from necessary for stability, unless phase is also
considered.)
As argued in~\cite{monotoneLSU}, Section 3, and reviewed below,
failure of the conditions often means that a ``pseudo-oscillation'' exists in
the system (provided that delays in the feedback loop are large enough),
in the rough sense that there are trajectories that ``look'' oscillatory if
observed under very noisy conditions and for finite time-intervals.
This begs the more interesting question of whether true periodic solutions
exist.  It turns out that some analogs of this converse result are known,
for certain low-dimensional systems, see~\cite{nussbaum-malletparet,ivanov}.
In the context of failure of the SGT, Enciso recently provided a converse
theorem for a class of cyclic systems, see~\cite{germancyclic}.
The Goldbeter model is far from being cyclic, however.
Theorem~\ref{trid-theo} in this paper proves the existence of oscillations for
a class of monotone tridiagonal systems under delayed negative feedback, and 
the theorem is then illustrated with the Goldbeter circadian model.

We first review the basic setup from~\cite{monotoneTAC}.

\section{I/O Monotone Systems, Characteristics, and Negative Feedback}

We consider an input/output system
\be{iosys}
\frac{dx}{dt}=f(x,u),\;\; y=h(x) \,,
\ee
in which states $x(t)$ evolve on some subset $X\subseteq \R^n$, and input and output
values $u(t)$ and $y(t)$ belong to subsets $U\subseteq \R^m$ and $Y\subseteq \R^p$ 
respectively. 
The maps $f:X\times U\rightarrow \R^n$ and $h:X\rightarrow Y$ are taken to be continuously
differentiable.
An \emph{input} is a signal $u:[0,\infty )\rightarrow U$ which is locally essentially compact
(meaning that images of restrictions to finite intervals are compact), and
we write $\varphi(t,x_0,u)$ for the solution of the initial value problem
$dx/dt(t)=f(x(t),u(t))$ with $x(0)=x_0$, or just $x(t)$ if $x_0$ and $u$ are
clear from the context, and $y(t)=h(x(t))$.
Given three partial orders on $X,U,Y$
(we use the same symbol $\prec$ for all three orders),
a \emph{monotone input/output system (``MIOS'')}, with
respect to these partial orders,
is a system~(\ref{iosys}) which is forward-complete
(for each input, solutions do not blow-up on finite time, so $x(t)$ and $y(t)$
are defined for all $t\geq 0$),
$h$ is a monotone map (it preserves order) and:
for all initial states $x_1,x_2$
for all inputs $u_1,u_2$,
the following property holds:
if $x_1$$\preceq$$x_2$
and $u_1$$\preceq$$u_2$
(meaning that 
$u_1(t)$$\preceq$$u_2(t)$ for all $t$$\geq $$0$), then
$\varphi(t,x_1,u)$$\preceq$$\varphi(t,x_2,u_2)$
for all $t>0$.
Here we consider partial orders induced by closed proper cones
$K\subseteq \R^\ell$, in the sense that $x\preceq y$ iff $y-x\in K$.
The cones $K$ are assumed to have a nonempty interior and are pointed, i.e.\
$K\bigcap -K=\{0\}$.
When there are no inputs nor outputs, the definition of monotone systems
reduces to the classical one of monotone dynamical systems studied by Hirsch,
Smith, and others \cite{Smith},
which have especially
nice dynamics.  Not only is chaotic or other irregular behavior ruled out,
but, in fact, under additional technical conditions (strong monotonicity)
almost all bounded trajectories converge to the set of steady states
(Hirsch's generic convergence theorem~\cite{Hirsch,Hirsch2}).

The most interesting particular case is that in which $K$ is 
an \emph{orthant} cone in $\R^n$, i.e.\ a set $S_\varepsilon $ of the form 
$\{x\in \R^n\,|\, \varepsilon _i x_i\geq 0\}$, where $\varepsilon _i=\pm 1$ for each $i$.
A useful test for monotonicity with respect to arbitrary orthant cones
(``Kamke's condition'' in the case of systems with no inputs and outputs) is
as follows.
Let us assume that all the
partial derivatives $\frac{\partial f_i}{\partial x_j}(x,u)$ for
$i\not= j$, $\frac{\partial f_i}{\partial u_j}(x,u)$ for all $i,j$, and 
$\frac{\partial h_i}{\partial x_j}(x)$ for all $i,j$
(subscripts indicate components)
do not change sign, i.e., they are either always $\geq 0$ or always $\leq 0$.
We also assume that $X$ is convex (much less is needed.)
We then associate a directed graph $G$ to the given MIOS, 
with $n+m+p$ nodes, and edges labeled ``$+$'' or ``$-$'' (or $\pm1$),
whose labels are determined by the signs of the appropriate partial
derivatives (ignoring diagonal elements of $\partial f/\partial x$).
One may define in an obvious manner undirected loops in $G$, and
the \emph{parity} of a loop is defined
by multiplication of signs along the loop.
(See e.g.~\cite{monotoneMulti} for more details.)
A system is monotone with respect to \emph{some}
orthant cones in $X,U,Y$ if and only if there are no negative loops in $G$.
In particular, if the cone is the main orthant ($\varepsilon =(1,\ldots ,1)$), the
requirement is that all partial derivatives must be nonnegative, with the
possible exception of the diagonal terms of the Jacobian of $f$ with respect
to $x$. 
A monotone system with respect to the main orthant is also called a
cooperative system.  
This condition can be extended to
non-orthant cones, see~\cite{%
Schneider_Vidyasagar,Volkman,Walcher,Walter}.

In order to define negative feedback (``inhibitory feedback'' in biology)
interconnections we will say that a system is \emph{anti-monotone} (with
respect to given orders on input and output value spaces) if the conditions
for monotonicity are satisfied, except that the output map \emph{reverses}
order: $x_1\preceq x_2 \Rightarrow h(x_2)\preceq h(x_1)$.

\subsection{Characteristics}

A useful technical condition that simplifies statements (one may weaken the
condition, see~\cite{malisoff-leenheer})
is that of existence of single-valued characteristics, which one may
also think of as
step-input steady-state responses or (nonlinear) DC gains.
To define characteristics, we consider the effect of a \emph{constant} input
$u(t)\equiv u_0$, $t\geq 0$ and study the
dynamical system $dx/dt=f(x,u_0)$.
We say that a single-valued characteristic exists if 
for each $u_0$ there is a state
$K(u_0)$ so that the system is globally attracted to $K(u_0)$, and in that
case we define the \emph{characteristic} $k:U\rightarrow Y$ as the composition
$h\circ K$.
It is remarkable fact for monotone systems that (under weak assumptions on
$X$, and boundedness of solutions) just knowing that a unique steady state
$K(u_0)$ exists, for a given input value $u_0$, already implies that $K(u_0)$
is in fact a globally asymptotically stable state for $dx/dt=f(x,u_0)$,
see~\cite{JiFa,dancer98}.

\subsection{Negative feedback}

Monotone systems with well-defined characteristics constitute useful building
blocks for arbitrary systems, and they behave in many senses like
one-dimensional systems.  Cascades of such systems inherit the same properties
(monotone, monostable response).  Under negative feedback, one obtains
non-monotone systems, but such feedback loops sometimes may be profitably
analyzed using MIOS tools.

\begin{figure}[ht]
\centering
\includegraphics[scale=0.25]{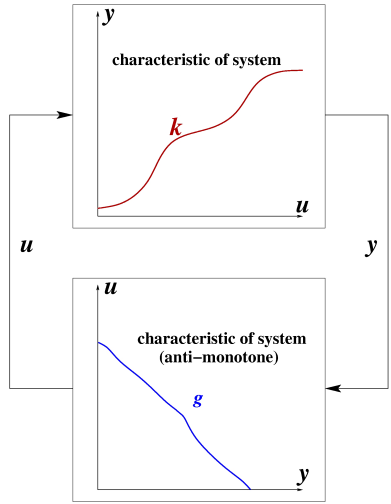}
\caption{Negative feedback configuration}
\label{oldfigs78a}
\end{figure}

We consider a feedback interconnection of a monotone and an anti-monotone
input/output system:
\beq
\label{sys1}
\frac{dx_1}{dt}&=&f_1(x_1,u_1),\;\; y_1=h_1(x_1)\\
\label{sys2}
\frac{dx_2}{dt}&=&f_2(x_2,u_2), \;\; y_2=h_2(x_2)
\eeq
with characteristics denoted by ``$k$'' and ``$g$'' respectively.
(We can also include the case when the second system is a static
function $y_2(t)=g(u_2(t))$.)
As in~\cite{monotoneMulti}, we will require here that the inputs and outputs
of both systems are scalar: $m_1$$=$$m_2$$=$$p_1$$=$$p_2$$=1$; the general
case~\cite{enciso_sontagSCL05} is similar but requires more notation and
is harder to interpret graphically.
The feedback interconnection of the systems~(\ref{sys1})
and~(\ref{sys2}) is obtained by letting
$u_2$$=$$y_1$$=$``$y$'' and $u_1$$=$$y_2$$=$``$u$'',
as depicted (assuming the usual real-number orders on inputs and outputs)
in Figure~\ref{oldfigs78a}.

\begin{figure}[ht]
\centering
\includegraphics[scale=0.25]{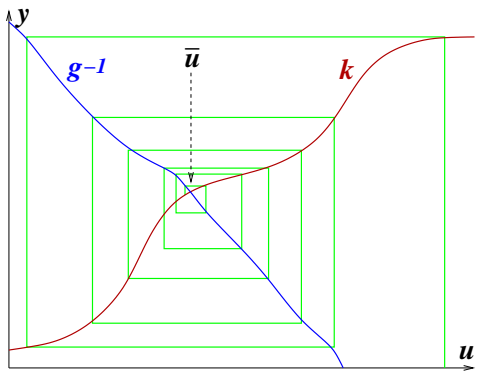}
\caption{characteristics}
\label{oldfigs78b}
\end{figure}

The main result from~\cite{monotoneTAC}, which we'll refer to as 
the monotone SGT theorem, is as follows.
We plot together $k$ and $g^{-1}$, as shown in 
Figure~\ref{oldfigs78b},
and consider the following discrete dynamical
system:
\[
u^+=(g\circ k)(u)
\]
on $U$.
Then, provided that solutions of the closed-loop system are bounded, the 
result is that, if
this iteration has a globally attractive fixed point 
${\bar  u}$, as shown in Figure~\ref{oldfigs78b} through a ``spiderweb''
diagram, then the feedback system has a globally attracting steady state.  
(An equivalent condition, see Lemma 2.3 in~\cite{predatorpreysgt05},
and~\cite{enciso-dcds}, is that 
the discrete system should have no nontrivial period-two orbits, i.e.
the equation $(g\circ k)(u)=u$ has a unique solution.)

It is not hard to prove, furthermore, that arbitrary delays may be allowed in
the feedback loop.  In other words, the feedback could be of the form $u(t) =
y(t-h)$, and such delays (even with $h=h(t)$ time varying or even
state-dependent, as long as $t-h(t)\rightarrow \infty $ as $t\rightarrow \infty $) do not destroy global
stability of the closed loop.  Moreover, it is also
known~\cite{enciso_smith_sontagJDE06} that diffusion does not destroy global
stability: a reaction-diffusion system, with Neumann boundary conditions,
whose reaction can be modeled in the shown feedback configuration, has the
property that all solutions converge to a (unique) uniform in space solution.

\subsection{Robustness}

It is important to point out that characteristics (called dose response
curves, activity plots, steady-state expression of a gene in response to an
external ligand, etc.) are frequently readily available from experimental
data, especially in molecular biology and pharmacology, in contrast to the
rare availability and high uncertainty regarding the precise form of the
differential equations defining the dynamics and values for all parameters
(kinetic constants, for example) appearing in the equations.
MIOS analysis allows one to combine the numerical information provided by
characteristics with the qualitative information given by ``signed network
topology'' (Kamke condition) in order to predict global behavior.
(See~\cite{ejc05es} for a longer discussion of this
``qualitative-quantitative approach'' to systems biology.)
The conclusions from applying the monotone SGT are robust with
respect to all perturbations that preserve monotonicity and stability
properties of the 1-d iteration.

Moreover, even if one would have a complete system specification,
the 1-d iteration plays a role vaguely analogous to that of Nyquist plots 
in classical control design, where the use of a simple plot allows quick
conclusions that would harder to obtain, and be far less intuitive, when
looking at the entire high-dimensional system.

\section{Existence of Characteristics}

The following result is useful when showing that characteristics exist for
some systems of biological interest, including the protein part of the
circadian model described later.  The constant $c$ represents the value
of a constant control $u(t)\equiv c$.

\bt{main-P-theo}
Consider a system of the following form:
\beqn
\dot x_0&=&c-\alpha _0(x_0)+\beta _0(x_1)\\
&\vdots&\\
\dot x_i&=&\alpha _{i-1}(x_{i-1})-\beta _{i-1}(x_i)-\alpha _i(x_i)+\beta _i(x_{i+1})\\
&& \quad\quad\quad\quad i=1,\ldots ,n-2\\
&\vdots&\\
\dot x_{n-1}&=&\alpha _{n-2}(x_{n-2})-\beta _{n-2}(x_{n-1}) -\alpha _{n-1}(x_{n-1})\\
&& \quad\quad\quad\quad +\beta _{n-1}(x_n) -\theta (x_{n-1})\\
\dot x_n&=&\alpha _{n-1}(x_{n-1})-\beta _{n-1}(x_n)
\eeqn
evolving on $\R_{\geq 0}^{n+1}$, where $c\geq 0$ is a constant.
Assume that $\theta $ and all the $\alpha _i, \beta _i$ are differentiable functions
$[0,\infty ) \rightarrow  [0,\infty )$ with everywhere positive derivatives
and vanishing at $0$,
\[
\theta  \;\;\mbox{and}\;\;  \alpha _i, \beta _i, \, i=0,\ldots ,n-2 \;\; \mbox{are bounded,}
\]
and 
\[
\alpha _{n-1}, \beta _{n-1} \;\; \mbox{are unbounded}.
\]
We use the notation $\theta (\infty )$ to indicate $\lim_{r\rightarrow \infty }\theta (r)$, and similarly
for the other bounded functions.
Furthermore, suppose that the following conditions hold:
\be{C2}
\alpha _{i-1}(\infty )+\beta _i(\infty ) < \alpha _i(\infty ) + \beta _{i-1}(\infty )\,\;\;
i=1,\ldots ,n-2
\ee
\be{C1}
\theta (\infty )+\beta _i(\infty )<\alpha _i(\infty ),\;\; i=0,\ldots ,n-2
\ee
\be{C3}
c < \theta (\infty ) .
\ee
Then, there is a (unique) globally asymptotically stable equilibrium for
the system.
\ets

Observe that~(\ref{C1}) (applied with $i=0$) together with~(\ref{C3}) imply
that also: 
\be{C1p}
c+\beta _0(\infty )<\alpha _0(\infty ) \,.
\ee

\bpr
We start by noticing that solutions are defined for all $t\geq 0$.
Indeed, consider any maximal solution $x(t)=(x_0(t),x_1(t),\ldots ,x_n(t))$.
From
\be{sumP}
\frac{d}{dt} \left( x_0 + x_1 +\ldots  +x_n\right) \;=\;
c - \theta (x_{n-1}) \;\leq \; c\,,
\ee
we conclude that there is an estimate
$x_i(t)\leq \sum_ix_i(t)\leq \sum_ix_i(0) + tc$
for each coordinate of $x$, and hence that there are no finite escape times.

Moreover, we claim that $x(\cdot )$ is bounded.
We first show that $x_0,\ldots ,x_{n-2}$ are bounded.
For $x_0$, it is enough to notice that
$\dot x_0\leq c-\alpha _0(x_0)+\beta _0(\infty )$, so that
\[
x_0(t)> \alpha _0^{-1}(c+\beta _0(\infty )) \;\;\Rightarrow \;\; \dot x_0(t)<0
\]
so~(\ref{C1p}) shows that $x_0$ is bounded.
Similarly, for $x_i$, $i=1,\ldots ,n-2$ we have that
\[
\dot x_i\leq \alpha _{i-1}(\infty )-\beta _{i-1}(x_i)-\alpha _i(x_i)+\beta _i(\infty )
\]
so~(\ref{C2}) provides boundedness of these coordinates as well.

Next we show boundedness of $x_{n-1}$ and $x_n$.

Since the system is a strongly monotone tridiagonal system, we know 
(see~\cite{Sm}, Corollary 1), that
$x_n(t)$ is {\em eventually monotone\/}.
That is, for some $T>0$, either
\be{mon1}
\dot x_n(t) \geq  0 \; \; \;\forall\,t\geq T
\ee
or
\be{mon2}
\dot x_n(t) \leq  0 \; \; \;\forall\,t\geq T \,.
\ee
Hence, $x_n(t)$ admits a limit, either finite or infinite.

Assume first that $x_n$ is unbounded, which means that
$x_n(t)\rightarrow \infty $ because of eventual monotonicity.
Then, case~(\ref{mon2}) cannot hold,
so~(\ref{mon1}) holds.
Therefore,
\[
\alpha _{n-1}(x_{n-1}(t))-\beta _{n-1}(x_n(t)) \;=\; \dot x_n \;\geq \;0
\]
for all $t\geq T$, which implies that
\[
x_{n-1}(t) \;\geq \; \alpha _{n-1}^{-1}(\beta _{n-1}(x_n(t)))\rightarrow \infty 
\]
as well.
Looking again at~(\ref{sumP}), and using that
$c-\theta (\infty )<0$ (property~(\ref{C3})), we conclude that
\[
\frac{d}{dt} \left( x_0 + x_1 + \ldots + x_{n-1} + x_n\right)(t)\;<\;0
\]
for all $t$ sufficiently large.
Thus $x_0 + x_1 + \ldots + x_{n-1} + x_n$ is bounded (and nonnegative), and this implies
that $x_{n-1}$ is bounded, a contradiction since we showed that $x_{n-1}\rightarrow \infty $.
So $x_n$ is bounded.

Next, notice that
$\dot x_{n-1}\leq \alpha _{n-2}(x_{n-2})+\beta _{n-1}(x_n)-\alpha _{n-1}(x_{n-1})$.
The two positive terms are bounded,
because both $x_{n-2}$ and $x_n$ are bounded.
Thus,
\[
\dot x_{n-1} \leq  v(t) - \alpha _{n-1}(x_{n-1}) \,,
\]
where $0\leq v(t)\leq k$ for some constant $k$.
Thus $\dot x_{n-1}(t)<0$ whenever $x_{n-1}(t)>\alpha _{n-1}^{-1}(k)$, and this proves that $x_{n-1}$
is bounded, as claimed.

Once that boundedness has been established, if we also show that there is a
unique equilibrium, then the theory of strongly monotone tridiagonal systems
(\cite{Sm,Smith}) (or~\cite{JiFa,dancer98} for more general monotone systems
results) will ensure global asymptotic stability of the equilibrium.  So we
show that equilibria exist and are unique.

Let us write $f_i(x)$ for the right-hand sides of the equations, so that
$\dot x_i=f_i(x)$ for each $i$.  We need to show that there is a unique
nonnegative solution $x=(x_0,\ldots ,x_n)$ of 
\[
f_0(x)=\ldots =f_n(x)=0.
\]
Equivalently, we can write the equations like this:
\beq
\label{eqn:fn}
f_n(x)&=& 0\\
\nonumber&\vdots&\\
\label{eqn:fi}
f_i(x)+\ldots +f_n(x) &=& 0\\
\nonumber&\vdots&\\
\label{eqn:allf}
f_0(x)+f_1(x)\ldots +f_n(x) &=& 0
\eeq
Since 
$f_0(x)+f_1(x)\ldots +f_n(x)=c-\theta (x_{n-1})$, (\ref{eqn:allf}) has the unique
solution $x_{n-1}=\bar x_{n-1}=\theta ^{-1}(c)$ 
which is well-defined because property~(\ref{C3}) says that $c < \theta (\infty )$.

Next, we consider equation~(\ref{eqn:fn}).
This equation has the unique solution:
\[
x_n \;=\; \bar x_n \;=\; \beta _{n-1}^{-1}(\alpha _{n-1}(\bar x_{n-1}))
\]
which is well-defined because $\beta _{n-1}$ is a bijection.

Pick $i\in \{1,\ldots ,n-1\}$ and
suppose that we have uniquely determined
$x_j=\bar x_j$ for each $j\geq i$.
We will show that $x_{i-1}$ is also uniquely defined.
Equation~(\ref{eqn:fi}) is:
\[
\alpha _{i-1}(x_{i-1})-\beta _{i-1}(\bar x_i)-\theta (\bar x_{n-1}) \;=\;0
\]
and has the unique solution
\[
x_{i-1} \;=\; \bar x_{i-1} \;=\;
\alpha _{i-1}^{-1}(\beta _{i-1}(\bar x_i)+\theta (\bar x_{n-1}))
\]
which is well-defined because property~(\ref{C1}) says that
$\theta (\infty )+\beta _{i-1}(\infty )<\alpha _{i-1}(\infty )$ for each $i=1,\ldots ,n-1$.
By induction on $i=n-1,\ldots ,1$, we have completed the uniqueness proof.
\epr

\section{The Goldbeter Circadian Model}

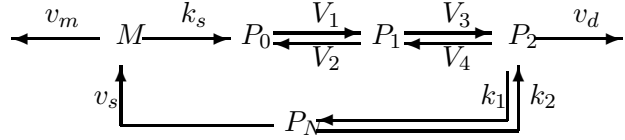
\begin{figure}[ht]
\begin{center}
\setlength{\unitlength}{1800sp}%
\begin{picture}(8424,1647)(2389,-5848)
\thicklines
\put(3601,-4561){\vector(-1, 0){1200}}
\put(4201,-4561){\vector( 1, 0){1200}}
\put(9601,-4561){\vector( 1, 0){1200}}
\put(6001,-4486){\vector( 1, 0){1200}}
\put(7801,-4486){\vector( 1, 0){1200}}
\put(7201,-4636){\vector(-1, 0){1200}}
\put(9001,-4636){\vector(-1, 0){1200}}
\put(6001,-5761){\line(-1, 0){2100}}
\put(3901,-5761){\vector( 0, 1){825}}
\put(6601,-5836){\line( 1, 0){2775}}
\put(9376,-5836){\vector( 0, 1){900}}
\put(9226,-5011){\line( 0,-1){675}}
\put(9226,-5686){\vector(-1, 0){2625}}
\put(2851,-4336){$v_m$}
\put(3826,-4636){$M$}
\put(7351,-4636){$P_1$}
\put(5551,-4636){$P_0$}
\put(9226,-4636){$P_2$}
\put(4726,-4336){$k_s$}
\put(8326,-4336){$V_3$}
\put(8326,-4936){$V_4$}
\put(6526,-4936){$V_2$}
\put(6526,-4336){$V_1$}
\put(10126,-4336){$v_d$}
\put(3526,-5461){$v_s$}
\put(6150,-5836){$P_N$}
\put(8851,-5461){$k_1$}
\put(9526,-5461){$k_2$}
\end{picture}
\caption{Goldbeter's Model}
\label{reactions}
\end{center}
\end{figure}

The original Goldbeter model of Drosophila circadian rhythms is 
schematically shown in Figure~\ref{reactions}.
The assumption is that PER protein is synthesized at a rate proportional to
its mRNA concentration.
Two phosphorylation sites are available, and constitutive
phosphorylation and dephosphorylation occur with saturation dynamics,
at maximum rates $v_i$ and with Michaelis constants $K_i$.
Doubly phosphorylated PER is degraded, also described by
saturation dynamics (with parameters $v_d, k_d$), and it is translocated 
to the nucleus, with rate constant $k_1$.
Nuclear PER inhibits transcription of the {\em per\/} gene,
with a Hill-type reaction of cooperativity degree $n$ and threshold constant
$K_I$.  The resulting mRNA is produced. and translocated to the cytoplasm, at
a rate determined by a constant $v_s$.  Additionally, there is saturated
degradation of mRNA (constants $v_m$ and $k_m$).

Corresponding to these assumptions, and assuming a well-mixed system, one
obtains an ODE system for concentrations are as follows:
\beq
\nonumber
\dot M &=& \frac{v_sK_I^n}{K_I^n + P_N^n}-\frac{v_mM}{k_m + M}\\
\nonumber
\dot P_0 &=& k_sM-\frac{V_1P_0}{K_1 + P_0}+\frac{V_2P_1}{K_2 + P_1}\\
\label{originalsys}
\dot P_1 &=& \frac{V_1P_0}{K_1 + P_0}-\frac{V_2P_1}{K_2 + P_1}
     -\frac{V_3P_1}{K_3 + P_1}+\frac{V_4P_2}{K_4 + P_2}\\ 
\nonumber
\dot P_2&=& \frac{V_3P_1}{K_3 + P_1}-\frac{V_4P_2}{K_4 + P_2}
   -k_1P_2+k_2P_N-\frac{v_dP_2}{k_d + P_2}\\
\nonumber
\dot P_N&=&k_1P_2-k_2P_N
\eeq
where the subscript $i=0,1,2$ in the concentration $P_i$ indicates the 
degree of phosphorylation of PER protein, $P_N$ is used to indicate the
concentration of PER in the nucleus, and $M$ indicates the concentration of
{\em per\/} mRNA.

The parameters (in suitable units $\mu M$ or $h^{-1}$)
used by Goldbeter are given in Table~\ref{table}.
\begin{table}[ht]
\begin{center}
\begin{tabular}{||c|c||c|c||}
\hline 
Parameter & Value & Parameter & Value \\
\hline
$k_2$ & 1.3 &
$k_1$ & 1.9 \\
$V_1$ & 3.2 &
$V_2$ & 1.58 \\ 
$V_3$ & 5 &
$V_4$ & 2.5 \\
$v_s$ & 0.76 &
$k_m$ & 0.5 \\
$k_s$ & 0.38 &
$v_d$ & 0.95 \\
$k_d$ & 0.2 &
$n$ & 4 \\
$K_1$ & 2 &
$K_2$ & 2 \\
$K_3$ & 2 &
$K_4$ & 2 \\
$K_I$ & 1 
& $v_m$ & 0.65\\
\hline
\end{tabular} 
\end{center}
\caption{Parameter Values}
\label{table}
\end{table}
With these parameters, there are limit cycle oscillations.
If we take $v_s$ as a bifurcation parameter, a Hopf bifurcation occurs
at $v_s\approx0.638$.

As an illustration of the SGT, we will show now that the theorem applies when
$v_s=0.4$.  This means that not only will stability of an equilibrium hold
globally in that case, but this stability will persist even if one introduces
delays to model the transcription or translation processes.  (Without loss of
generality, we may lump these delays into one delay, say in the term $P_N$
appearing in the equation for $M$.)
On the other hand, we'll see later that the SGT discrete iteration does not
converge, and in fact has a period-two oscillation, when $v_s=0.5$.
This suggests that periodic orbits exist in that case, at least if
sufficiently large delays are present, and we analyze the existence of such
oscillations.

For the theoretical developments, we assume from now on that
\be{vs}
v_s \leq  0.54
\ee
and the remaining parameters will be constrained below, in such a manner that
those in the previously given table will satisfy all the constraints.

\subsection{Breaking-up the Circadian System and Applying the SGT}

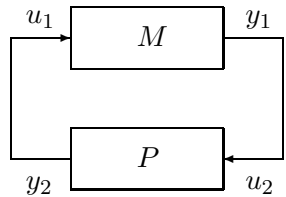
\begin{figure}[ht]
\begin{center}
\setlength{\unitlength}{2500sp}%
\begin{picture}(2724,1866)(3889,-3715)
\put(4051,-2011){$u_1$}
\put(4051,-3661){$y_2$}
\put(6226,-2011){$y_1$}
\put(6226,-3661){$u_2$}
\thinlines
\put(4501,-2461){\framebox(1500,600){}}
\put(6001,-2161){\line( 1, 0){600}}
\put(6601,-2161){\line( 0,-1){1200}}
\put(6601,-3361){\vector(-1, 0){600}}
\put(4501,-3361){\line(-1, 0){600}}
\put(3901,-3361){\line( 0, 1){1200}}
\put(3901,-2161){\vector( 1, 0){600}}
\put(4501,-3661){\framebox(1500,600){}}
\put(5150,-2236){$M$}
\put(5150,-3436){$P$}
\end{picture}
\caption{Systems in feedback}
\label{interconnection}
\end{center}
\end{figure}

We choose to view the system as the feedback interconnection
of two subsystems, one for $M$ and the other one for $P$,
see Figure~\ref{interconnection}.

\subsubsection*{mRNA Subsystem}

The mRNA ($M$) subsystem is described by the scalar differential equation
\[
\dot M \;= \; \frac{v_sK_I^n}{K_I^n + u_1^n}-\frac{v_mM}{k_m + M}
\]
with input $u_1$ and output $y_1=k_sM$.

As state-space, we will pick a compact interval
$X_1=[0,\bar M]$, where
\be{ineqM}
\frac{v_sk_m}{v_m-v_s} \leq  \bar M < \frac{v_d}{k_s}
\ee
and we assume that $v_s<v_m$.
The order on $X_1$ is taken to be the usual order from $\R$.

Note that the first inequality implies that
\be{eqn1}
v_s < \frac{v_m \barM}{k_m+\barM}
\ee
and therefore
\[
\frac{v_sK_I^n}{K_I^n + u_1^n}-\frac{v_m\barM}{k_m + \barM} < 0
\]
for all $u_1\geq 0$,
so that indeed $X_1$ is forward-invariant for the dynamics.

With the parameters shown in the table given earlier  (except for $v_s$, which
is picked as in~(\ref{vs})),
\[
\barM = 2.45
\]
satisfies all the constraints.

As input space for the mRNA system, we pick $U_1=\R_{\geq 0}$, and as output space
 $Y_1=[0,v_d)$.
 Note that $y_1 = k_s M \leq  k_s \barM < v_d$, by~(\ref{ineqM}), so the output
belongs to $Y_1$.
We view $U_1$ as having the \emph{reverse} of the usual order, and $Y_1$ are
is given the usual order from $\R$. 

The mRNA system is monotone, because it is internally monotone
($\partial f/\partial u<0$, as required by the reverse order on $U_1$) and the
output map is monotone as well.

Existence of characteristics is immediate from the fact that
$\dot M>0$ for $M<k(u_1)$ and $\dot M<0$ for $M>k(u_1)$, where, for each
constant input $u_1$, 
\[
k(u_1)=
\frac {
v_s\,K_I^n\,k_m
}{
v_m\,K_I^n+v_m\,{u_1}^{n}-v_s\,K_I^n
}
\]
(which is an element of $X_1$).

Note that all solutions of the differential equations which describe the
$M$-system, even those that do not start in $X_1$, enter $X_1$ in finite
time (because $\dot M(t)<0$ whenever $M(t)\geq \barM$, for any input $u_1(\cdot )$).
The restriction to the state space $X_1$ (instead of using all of $\R_{\geq 0}$)
is done for convenience, so that one can view the output of the $M$ system as
in input to the $P$-subsystem.  (Desirable properties of the $P$-subsystem
depend on the restriction imposed on $U_2$.)  Given any trajectory, its
asymptotic behavior is independent of the behavior in an initial finite time
interval, so this does not change the conclusions to be drawn.  (Note that
solutions are defined for all times --no finite explosion times-- because the
right-hand sides of the equations have linear growth.)

\subsubsection*{Protein Subsystem}

The second ($P$) subsystem is four-dimensional:
\beqn
\dot P_0 &=& u_2-\frac{V_1P_0}{K_1 + P_0}+\frac{V_2P_1}{K_2 + P_1}\\
\dot P_1 &=& \frac{V_1P_0}{K_1 + P_0}-\frac{V_2P_1}{K_2 + P_1}
  -\frac{V_3P_1}{K_3 + P_1}+\frac{V_4P_2}{K_4 + P_2}\\
\dot P_2&=& \frac{V_3P_1}{K_3 + P_1}-\frac{V_4P_2}{K_4 + P_2}
  -k_1P_2+k_2P_N-\frac{v_dP_2}{k_d + P_2}\\
\dot P_N&=&k_1P_2-k_2P_N
\eeqn
with input $u_2$ and output $y_2=P_N$.

For the $P$ subsystem, the state space is $\R_{\geq 0}^4$ with the main orthant
order, and the input space is $U_2=Y_1$ and the output space is $Y_2=U_1$
(with the orders specified earlier).
Internal monotonicity of the $P$ subsystem is clear from the fact that
$\frac{\partial \dot P_i}{\partial P_j}>0$ for all $i\not= j$ (cooperativity).
In fact, because these inequalities are strict and the Jacobian matrix is
tridiagonal and irreducible at every point, this is an example of a 
{\em strongly monotone tridiagonal system\/}
(\cite{Sm,Smith}).
The system is anti-monotone because the identity mapping reverses order
(recall that $Y_2=U_1$ has the reverse order, by definition).

We obtain the following result as a corollary of Theorem~\ref{main-P-theo},
applied with $n=3$, $\theta (r)=\frac{v_dr}{k_d + r}$, 
$\alpha _0(r)=\frac{V_1r}{K_1 + r}$, etc.
It says that, for the parameters in Table~\ref{table}, as well as for a
larger set of parameters, the system has a well-defined characteristic, which
we will denote by $g$.
(It is possible to give an explicit formula for $g^{-1}$, in this example.)

\bp{main-P-prop}
Suppose that the following conditions hold:
\bi
\item
$v_d + V_2 < V_1$
\item
$V_1 + V_4 < V_2 + V_3$
\item
$0\leq c < v_d$
\item
$V_4+v_d<V3$
\ei
and that all constants are positive and the input $u_2(t)\equiv c$.
Then the $P$-system has a unique globally asymptotically stable equilibrium.
\eps

\section{Closing the Loop}

Solutions of the closed-loop system, i.e., of the original
system~(\ref{originalsys}), are bounded under the above assumptions.
To see this, we argue as follows.
Take any solution of the closed loop system.
As we pointed out earlier, there are no finite time explosions, and also
the $M$ coordinate will converge to the set $X_1=[0,\bar M]$.

This means that the subsystem corresponding to the $P$-coordinates will be
forced by an input $u_2$ such that $u(t) \in [0,k_s \bar M]$ for all $t\geq
t_0$, for some $t_0$.
Now, for constant inputs in $[0,v_d)$, which contains $[0,k_s \bar M]$, we
have proved that a characteristic $k$ exists for the open-loop system
corresponding to these coordinates.
Therefore, by monotonicity, the trajectory components
$y(t)=(P_0(t),P_1(t),P_2(t),P_N(t))$ will lie in the main orthant order
rectangle $[y_0(t),y_1(t)]$, for each $t\geq0$,
where $y_0$ is the solution with constant input $u_2=0$ and $y_0(t_0)=y(t_0)$
and where $y_N$ is the solution with constant input $u_2=k_s \bar M$, and
$y_1(t_0)=y(t_0)$.
Since $y_0$ and $y_1$ converge to $[k(0),k(k_s \bar M)]$, the omega-limit set
of $y$ is included in $[k(0),k(k_s \bar M)]$, and therefore the $P$ components
are bounded as well.

\begin{figure}[ht]
\begin{center}
\includegraphics[scale=0.2]{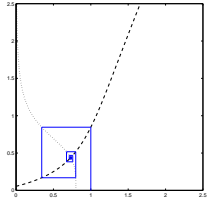}
\caption{Stability of spiderweb ($v_s=0.4$)}
\label{fig-circ4}
\end{center}
\end{figure}

Now we are ready to apply the main theorem in~\cite{monotoneTAC}.
In order to do this, we first need to plot the characteristics.
See Figure~\ref{fig-circ4} for the plots of $g$ and $k^{-1}$
(dashed and dotted curves) and the a typical ``spiderweb diagram'' 
(solid lines), when we pick the parameter $v_s=0.4$.
It is evident that there is global convergence of the discrete iteration.
Hence no oscillations can arise, even under
arbitrary delays in the feedback from $P_N$ to $M$, and in fact that all
solutions converge to a unique equilibrium.

\begin{figure}[ht]
\begin{center}
\includegraphics[scale=0.2]{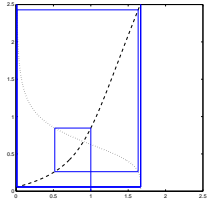}
\caption{Instability of spiderweb ($v_s=0.5$)}
\label{fig-circ5}
\end{center}
\end{figure}

On the other hand, for a larger value of $v_s$, such as $v_s=0.5$, the
discrete iteration conditions are violated;
see Figure~\ref{fig-circ5} for the ``spiderweb diagram'' that shows
divergence of the discrete iteration.
Thus, one may
expect periodic orbits in this case.
We next prove a result that shows that indeed that happens.

\section{Periodic Behavior when SGT Conditions Fail}

One may conjecture that there is a connection between periodic behaviors of
the original system, at least under delayed feedback, and of the associated
discrete iteration.  We first present an informal discussion and then give a
precise result.

For simplicity, let us suppose that $k$ already denotes the composition of
the characteristics $g$ and $k$.
The input values $u$ with $k(k(u))=u$ which do not arise from the
unique fixed point of $k(u)=u$ are period-two orbits of the iteration
$u^+=k(u)$.  Now suppose that we consider the delay differential system
$\dot x(t)=f(x(t),h(x(t-r)))$, where the delay $r>0$ is very large.
We take the initial condition $x(t)=x_0$, $t\in [-r,0]$, where $x_0$ is picked
in such a manner that $h(x_0)=u_0$, and $u_0\not= u_1$ are two elements of $U$
such that $k(u_0)=u_1$ and $k(u_1)=u_0$.
If the input to the open-loop system $\dot x=f(x,u)$ is $u(t)\equiv u_0$, then the
definition of characteristic says that the solution $x(t)$ approaches $x_1$, 
where $h(x_1)=u_1$, 
Thus, if the delay length $r$ is large enough, the solution of the closed-loop
system will be close to the constant value $x_1$ for $t\approx r$.
Repeating this procedure, one can show the existence of a lightly damped
``oscillation'' between the values $x_0$ and $x_1$, in the sense of a
trajectory that comes close to these values as many times as desired
(a larger $r$ being in principle required in order to come closer and more
often). 
In applications in which measurements have poor resolution and time duration,
it may well be impossible to practically determine the difference
between such pseudo-oscillations and true oscillations.

It is an open question to prove the existence of true periodic orbits,
for large enough delays, when the small-gain condition fails.  The problem is
closely related to questions of singular perturbations for delay systems, by
time-re-parametrization.  We illustrate this relation by considering
the scalar case, and with $y=x$.
The system $\dot x=f(x,x(t-r))$, has periodic orbits for large enough $r$ if and
only if the system $\varepsilon \dot x(t)=f(x(t),x(t-1))$ has periodic orbits for
small enough $\varepsilon >0$.  For $\varepsilon =0$, we have the algebraic equation $f(x,u)=0$
that defines the characteristic $x=k(u)$.  Thus one would want to know that
periodic orbits of the iteration $u^+=k(u)$, seen as the degenerate case
$\varepsilon =0$, survive for small $\varepsilon >0$.
A variant of this statement is known in dimension one from work of Nussbaum
and Mallet-Paret
(\cite{nussbaum-malletparet}), which shows the existence of a continuum of
periodic orbits which arise in a Hopf bifurcation and persist for $0<\varepsilon \ll 1$;
see also the more recent work~\cite{ivanov}.
(We thank Hal Smith for this observation.)

We now show that, at least, for a class of systems which is of some general
interest in biology, and which contains the circadian model, oscillations can
be proved to exist if delays are large enough and the SGT fails locally
(exponential instability of the discrete iteration).

\subsection{Predicting Periodic Orbits when the Condition Fails}

In this section we prove the following theorem, which applies immediately to
the complete circadian model~(\ref{originalsys}).

\bt{trid-theo}
Consider a tridiagonal system $\dot x=f(x,u)$ with scalar input $u$:
\be{tridiagonal}
\begin{array}{rcl}
\dot {x}_1 &=& f_1 ( x_1,x_2,u)\\
\dot {x}_2 & = & f_2 (x_1,x_2,x_3) \\
&\vdots&\\
\dot {x}_{n-1} &=& f_{n-1} ( x_{n-2},x_{n-1},x_n ) \\
\dot {x}_n & = & f_n ( x_{n-1}, x_n )
\end{array}
\ee
and scalar output $y=x_n$.
The functions $f_i$ are twice continuously differentiable, and
(cooperativity) all the off-diagonal Jacobian entries are positive.
Suppose that there is a unique pair $(x_0,u_0)\in \R^n\times \R$ such that
$f(x_0,u_0)=0$, and consider the linearized system $(A,b,c)$, where
$b=(1,0,\ldots ,0,0)'$, $c=(0,0,\ldots ,0,1)$, and $A = D_xf$, the Jacobian of the
vector field $f(x,u_0)$ evaluated at $x_0$.
Assume that $A$ is 
nonsingular, 
and
let $g=c(-A)^{-1}b$ be the DC gain of the linearized system.
Pick any positive number $k$ such that $kg>1$, and consider a delayed feedback 
$u(t)= - k y(t- h)$ with $h>0$.
Then, for some $h>0$, the system~(\ref{tridiagonal}) under 
the feedback $u(t)= - k y(t- h_0)$ admits a periodic solution, and, moreover,
the omega-limit set of every bounded solution is either a periodic orbit,
the origin, or a nontrivial homoclinic orbit with $\lim_{t\rightarrow \pm\infty }=x_0$.
\ets

\medskip

Note that the uniqueness result for closed-loop equilibria will always hold in
our case, and the DC gain property $kg>1$ corresponds to a locally unstable
discrete iteration.  The matrix $A$ is Hurwitz when we have hyperbolicity and
parameters as considered earlier (existence of characteristics).  The
conclusion is that, for a suitable delay length $h_0$, there is at least one
periodic orbit, and, moreover, bounded solutions not converging to zero
exhibit oscillatory behavior (with periods possibly increasing to
infinity, if the omega-limit set is a homoclinic orbit).
(We conjecture that moreover, for the circadian example, in fact almost all
solutions converge to a periodic orbit.  Proving this would require
establishing that no homoclinic orbits exist for our systems, just as shown,
when no delays present, in~\cite{MPSmith}.)

Before proving Theorem~\ref{trid-theo}, we show the following simple lemma
about linear systems. 

\bl{linear-tri}
Consider a linear $n$-dimensional single-input single-output system $(A,b,c)$,
with $b=(1,0,\ldots ,0,0)'$ and $c=(0,0,\ldots ,0,1)$, and
suppose that $A$ is a linear tridiagonal matrix
\[
A \;=\;
\pmatrix{
d_1 & b_2 & 0   & 0   & \ldots  & 0 & 0 \cr
a_2 & d_2 & b_3 & 0   & \ldots  & 0 & 0 \cr
0   & a_3 & d_3 & b_4 & \ldots  & 0 & 0 \cr
\cdot   & \cdot  & \cdot  & \cdot  & \ldots  & \cdot  & \cdot  \cr
\cdot   & \cdot  & \cdot  & \cdot  & \ldots  & \cdot  & \cdot  \cr
0   & 0   & 0   & 0   & \ldots  & a_n & d_n
}
\]
with $a_ib_i>0$ for all $i$ (in particular, this holds if all
off-diagonal elements are positive).
Then, the transfer function $W(s)=c(sI-A)^{-1}b$ has no zeroes and
has distinct real poles; more specifically,
$W(s)= \frac{p_0}{q(s)}$, where $p_0=a_2\ldots a_n$ and
$q(s)=(s-\alpha _1)\ldots (s-\alpha _n)$ for distinct real numbers $\alpha _1,\ldots ,\alpha _n$.
Moreover, there are two real-valued functions $\mu :\C\rightarrow \R$ and $\nu :\C\rightarrow \R_{>0}$
so that the logarithmic derivative $Q(s)=q'(s)/q(s)$ satisfies
$Q(s)=\mu (s) - i\nu (s){\mbox{Im}s}$ for every $s$ that is not a root
of $q$.
\els

\bpr
The fact that $A$ has $n$ distinct real eigenvalues is a classical one in
linear algebra; we include a short proof to make the paper more self-contained.
Pick any positive number $\sigma _1$ and define, inductively,
\[
\sigma _i \,:= \; \sigma _{i-1} \sqrt{\frac{a_i}{b_i}}
\]
for $i=2,\ldots ,n$.
Let $S = \mbox{diag}\,\left(\sigma _1,\ldots ,\sigma _n\right)$.
Then $B = S^{-1}AS$ is a tridiagonal {\em symmetric} matrix:
\[
B \;=\;
\pmatrix{
d_1 & c_2 & 0   & 0   & \ldots  & 0 & 0 \cr
c_2 & d_2 & c_3 & 0   & \ldots  & 0 & 0 \cr
0   & c_3 & d_3 & c_4 & \ldots  & 0 & 0 \cr
\cdot   & \cdot  & \cdot  & \cdot  & \ldots  & \cdot  & \cdot  \cr
\cdot   & \cdot  & \cdot  & \cdot  & \ldots  & \cdot  & \cdot  \cr
0   & 0   & 0   & 0   & \ldots  & c_n & d_n
}
\]
where $c_i = \varepsilon _i\sqrt{a_ib_i}$
and $\varepsilon _i=\mbox{sign}\,a_i=\mbox{sign}\,b_i\in \{-1,+1\}$.
Therefore $B$, and hence also $A$, has all its eigenvalues real.
Moreover, there is a basis $\{v_1,\ldots ,v_n\}$ consisting of orthogonal
eigenvectors of $B$, and so $A$ admits the linearly independent eigenvectors
$Sv_i$.
Moreover, all eigenvalues of $B$ (and so of $A$) are distinct.
(Pick any $\lambda $ and consider $C:=B-\lambda I$.
The first $n-1$ rows of $C$ look just like those of $B$, 
with $d_i:=d_i-\lambda $.
The $n-1\times n$ matrix consisting of these rows has rank $n-1$ (just consider its
last $n-1$ columns, a nonsingular matrix), so it follows that $C$ has
rank$\geq n-1$.  Therefore, the kernel of $C$ has dimension at most one.)
We conclude that $A$ has $n$ distinct real eigenvalues 
and hence its characteristic polynomial has the
form $q(s)=(s-\alpha _1)\ldots (s-\alpha _n)$.

By Cramer's rule, $(sI-A)^{-1}=(1/q(s))\mbox{cof}(sI-A)$, where ``cof''
indicates matrix of cofactors.  Thus $W(s)=p_0/q(s)$, where $p_0$ is the
$(n,1)$ entry of $\mbox{cof}(sI-A)$, i.e.\ $(-1)^{n+1}$ times
the determinant of the matrix $\widetilde{(sI-A)}_{1,n}$ obtained by deleting
from $sI-A$ the first row and last column.  The matrix
$\widetilde{(sI-A)}_{1,n}$ is upper triangular, and its determinant is
$(-a_2)\ldots (-a_n) = (-1)^{n-1}a_2\ldots a_n$.
Therefore $p_0=a_2\ldots a_n$, as claimed.

Finally, consider $Q(s) = \sum_{k=1}^n \frac{1}{s+\alpha _k}$.
Write $s=a+ib$, so that
\[
\frac{1}{s+\alpha _k}
=\frac{1}{(a+\alpha _k) + ib}
=\frac{(a+\alpha _k)-ib}{\rho _k}
\]
where $\rho _k=(a+\alpha _k)^2+b^2$, and therefore
\[
Q(s) \;= \;
\sum_{k=1}^n \frac{a+\alpha _k}{\rho _k}
\,-\,
i \left(\sum_{k=1}^n \frac{1}{\rho _k}\right) b
\;=\;
\mu (s) - i\nu (s)b
\]
as desired.
\epr

We now continue the proof of Theorem~\ref{trid-theo}, by first studying the
closed-loop linearized system $\dot x(t) = Ax(t) + bkc x(t-h)$.
The closed-loop transfer function 
\[
W_h = \frac{W}{1 + ke^{-hs}W(s)}
\]
corresponding to a negative feedback loop with delay $h$ and gain 
$k$, simplifies to:
\[
W_h = \frac{p_0}{F}\,,\quad\quad F(s,h) = q(s) +  pe^{-hs},
\]
where $p=p_0k$.

In order to prove that there are oscillatory solutions for some
$h=h_0$, we proceed as follows.  We will use the weak form of the Hopf
bifurcation theorem (``weak'' in that no assertions are made regarding super or
subcriticality of the bifurcation) as given in Theorem 11.1.1 in~\cite{hale}.
The theorem guarantees that oscillatory solutions will exist, for the
nonlinear system, and for some value of the delay $h$ arbitrarily close to
a given $h_0>0$, provided that the following two properties hold for $h_0$:

\noindent{\bf(H1)}
There is some $\omega _0\not= 0$ such that  $F(i\omega _0,h_0)=0$,
$\omega =i\omega _0$ is a simple root of $F(\omega ,h_0)=0$, and
(nonresonance) $F(mi\omega ,h_0)\not= 0$ for all integers $m>1$;

\noindent
and letting $\lambda (h)$ be a ${\cal C}^1$ function such that $F(\lambda (h),h)=0$
for all $h$ near $h_0$ and $\lambda (h_0)=\omega _0$ (such a function always exists):

\noindent{\bf(H2)}
$\mbox{Re}\,\lambda '(h_0)\not= 0$.

In order to prove these properties, we proceed analogously to what is done for
cyclic systems in~\cite{germancyclic}.
(Cyclic systems are the special case in which
$\partial f_i/\partial x_{i+1}\equiv 0$ for each $i=1,\ldots ,n-1$, which is not the
case in our circadian system.)

We first show that $F(i\omega _0,h_0)=0$ for some $h_0>0$ and $\omega _0>0$.
Since $q(s)F(s,h) = 1 +  (p/q(s))e^{-hs}$ and
$q(i\omega )\not= 0$ for all real numbers $\omega $ (because $q$ has only real roots,
and $A$ is nonsingular, so also $q(0)\not= 0$),
it is enough to find an $h_0>0$ and $\omega _0>0$ such that
$f(\omega _0)=-e^{ih_0\omega _0}=e^{i(h_0\omega _0-\pi )}$, where $f(\omega )=p/q(i\omega )$.
Since $f$ is a continuous function on $[0,\infty )$,
$f(0)=p_0k/q(0) = W(0)k = gk>1$ by assumption,
and $\lim_{\omega \rightarrow \infty }f(\omega ) = 0$, there is some
$\omega _0>0$ such that $|f(\omega )|=1$, so that $f(\omega )=e^{i\varphi}$ for some $\varphi$
which we may take in the interval $(0,2\pi ]$.
It thus suffices to pick $h_0 = \varphi+\pi /\omega _0$, so that
$h_0\omega _0-\pi =\varphi$.

Fix any such $h_0$.
Since for retarded delay equations there are at most a finite number of
roots on any vertical line, we can pick $\omega _0$ with largest possible
magnitude, so that $F(mi\omega _0,h_0)\not= 0$ for all integers $m>1$.
To prove that (H1) and (H2) hold for these $h_0$ and $\omega _0$,
we first prove that $(dF/ds)(i\omega ,h)$ is nonzero at each point $i\omega $ in the
imaginary axis, and all $h>0$ (so, in particular, at $(i\omega _0,h_0)$).
By the implicit function theorem, this will imply that $\omega _0$ is a simple
root, as needed for (H1).
Since $F(\lambda (h),h))\equiv 0$ in a neighborhood of $h_0$, we also have that
\beqn
\lambda '(h) 
&=&
-
\frac{\partial F}{\partial h}(\lambda (s),h)
\,/\,
\frac{\partial F}{\partial s} (\lambda (s),h)\\
&=&
\frac{spe^{-hs}}{q'(s)-hpe^{-hs}}
\;=\;
\frac{s}{\frac{q'(s)}{pe^{-hs}}-h}
\,.
\eeqn
At $\omega =\omega _0$ and $h=h_0$, we have that
$F(\omega ,h)=0$, so $q(s)=-pe^{-hs}$, and therefore, denoting $Q(s):=q'(s)/q(s)$:
\[
\lambda '(h_0) 
\;=\;
-
\frac{i\omega _0}{Q(i\omega _0)+h_0}
\,.
\]
It follows that, for $\omega _0>0$:
\beqn
\mbox{Re}\,\lambda '(h_0)
&=&
\mbox{Im}\,\frac{1}{Q(i\omega )+h_0}\\
&=&
-
\mbox{Im}\,\left[Q(i\omega _0)+h_0\right]
\;=\;
-
\mbox{Im}\,Q(i\omega _0)
\,.
\eeqn
The proof that periodic orbits exist, for each $h$ near enough $h_0$,
will then be completed if we show that
$\mbox{Im}\,Q(i\omega )\not= 0$
for all nonzero real numbers $\omega $.
But, indeed, Lemma~\ref{linear-tri} says that
$Q(i\omega )=\mu (i\omega ) - i\nu (i\omega )\omega $,
where $\nu (i\omega )>0$, so $\mbox{Im}\,Q(i\omega )= -\nu (i\omega )\omega \not= 0$.

To conclude the proof, we note that the conclusion about global behavior
follows from the Poincar\'e-Bendixson for delay-differential tridiagonal
systems due to Mallet-Paret and Sell \cite{malletparet}.
\qed

Note that, since $\mbox{Re}\,\lambda '(h_0)\not= 0$, if $h_0'$ is near enough
$h_0$, then the system~(\ref{tridiagonal}) under negative feedback
$u=-ky(t-h_0')$ admits a pair of complex conjugate eigenvalues $a+i\omega $
for its linearization, with $a>0$.  Thus, its equilibrium is exponentially
unstable, and therefore every bounded solution not starting from the
center-stable manifold will in fact converge to either a homoclinic orbit
involving the origin or a periodic orbit.

\subsection{Examples}

As a first example, we take the system with the parameters that we have
considered, and $v_s=0.5$.  We have seen that the spider-web diagram suggests
oscillatory behavior when delays are present in the feedback loop.
We first compute the equilibrium of the closed-loop system (with no delay),
which is approximately:
\[
M \approx 1.47,
P_1 = 0.42,
P_2 = 0.29,
P_0 = 0.71,
P_N = 0.42.
\]
We now consider the system with variables $M,P_1,P_2,P_3,P_N$ in which the
feedback term $v_sK_I^n/(K_I^n+P_3^n)$ is replaced by an input $u$.
Let $A$ be the Jacobian of this open-loop dynamics evaluated at the positive
equilibrium given above.
Then $sI-A\approx$
\[
\left[ \begin {array}{ccccc}  0.08+s&0&0&0&0
\\\noalign{\medskip}- 0.38& 0.87+s&- 0.54&0&0
\\\noalign{\medskip}0&- 0.87& 2.24+s&- 0.96&0
\\\noalign{\medskip}0&0&- 1.70& 3.66+s&- 1.3
\\\noalign{\medskip}0&0&0&- 1.9&1.3+s\end {array} \right] 
\]
and hence the transfer function $W(s)=c(sI-A)^{-1}b$, where
$b=(1,0,0,0,0)'$ and $c=(0,0,0,0,1)$, is:
\[
W(s) =  \frac{p_0}{q(s)},
\]
where $p_0\approx1.075$ and $q(s)\approx$
\[
\left(0.084+s \right)  
\left(
{s}^4
+ 8.08\,{s}^{3}
+ 17.61\,{s}^{2}
+ 10.98\,s
+ 1.56
\right) 
.
\]
The DC gain of the system is $g=W(0)\approx8.26$ (which is positive, as it
should be since the open loop system is monotone and has a well-defined
steady-state characteristic).  and $k=-\partial
(v_sK_I^n/(K_I^n+P_3^n)/\partial P_3\approx
0.14$
 when evaluated at the
computed equilibrium.
Thus $f(0)=gk\approx1.138>1$, as required.
Indeed, $\mbox{Im}\,Q(i\omega)\approx$
\[
\frac {-
\left(
2.88
+ 133.26\,\omega ^{2}
+ 408.07\,\omega ^{4}
+ 120.12\,\omega ^{6}
+ 5.0\,\omega ^{8} 
\right)
\omega 
}
{
\left(
0.007+\omega ^{2}
\right) 
\left(  
2.42
+ 65.68\omega ^{2}
+ 135.81\omega ^{4}
+ 30.02\omega ^{6}
+\omega ^{8}
\right) 
}
\]
and hence
$\mbox{Im}\,Q(i\omega)\not= 0$
for all $\omega\not= 0$.

\begin{figure}[ht]
\begin{center}
\includegraphics[scale=0.4]{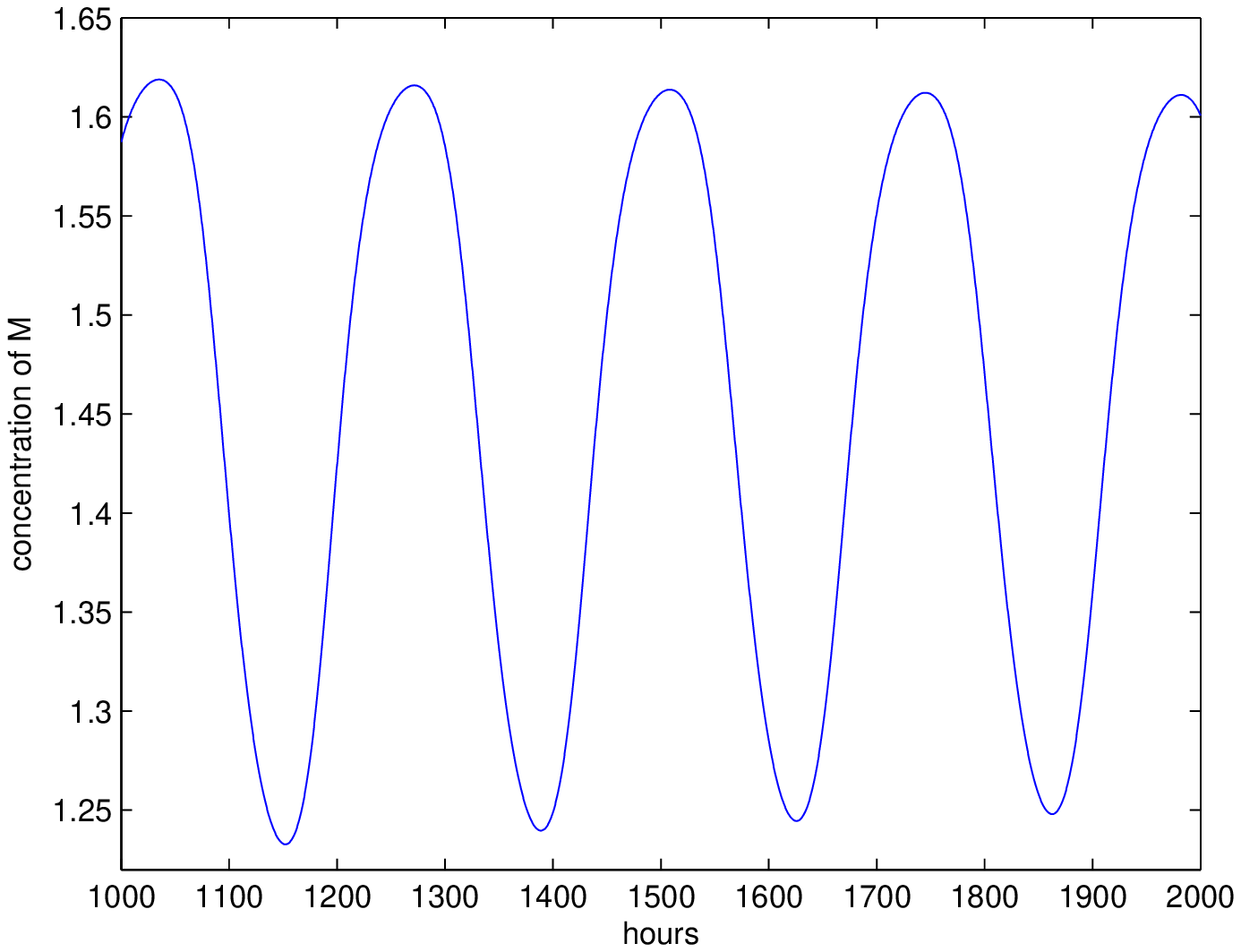}
\caption{Oscillations seen in simulations ($v_s=0.5$, delay $=100$hr, initial
conditions all at $0.2$), using MATLAB's dde23 package}
\label{fig-circ5-simul}
\end{center}
\end{figure}

\begin{figure}[ht]
\begin{center}
\includegraphics[scale=0.4]{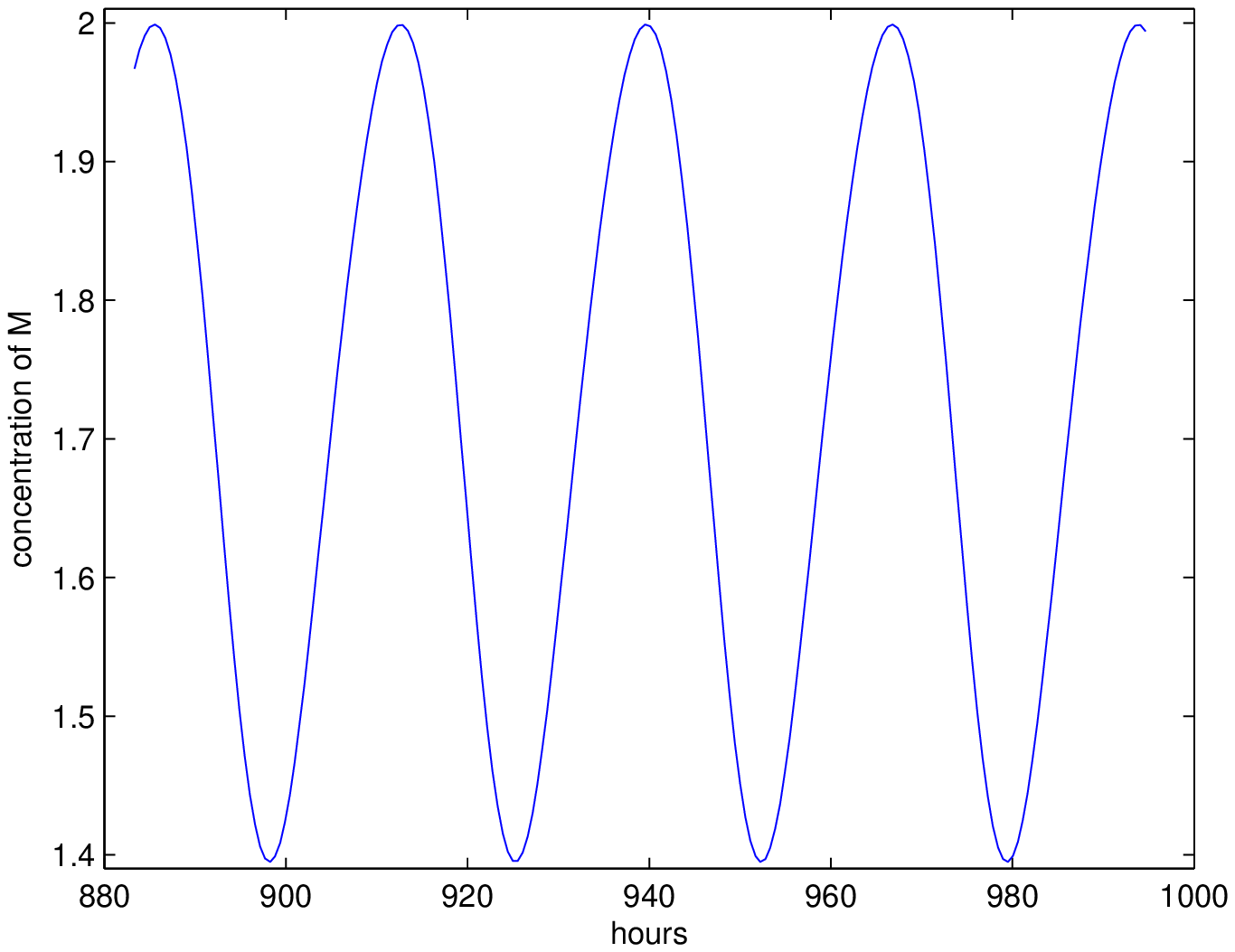}
\caption{Oscillations seen in simulations ($v_s=0.6$, delay $=1$hr, initial
conditions all at $0.2$), using MATLAB's dde23 package}
\label{fig-circ6-simul}
\end{center}
\end{figure}

We show in Figure~\ref{fig-circ5-simul} one simulation, with $h=100$, showing
a periodic limit cycle.
The delay length needed for oscillations when $v_s=0.5$ is biologically
unrealistic, so we also show simulations for $v_2=0.6$, a value for which 
no oscillations occur without delays, but for which oscillations (with a
period of about 27 hours) occur when the delay length is about 1 hour,
see Figure~\ref{fig-circ6-simul}.

\section{A Counterexample}

We now provide a (non-monotone) system as well as a feedback law $u=g(y)$ so
that: 
\bi
\item
the system has a well-defined and increasing characteristic $k$,
\item
the discrete iteration $u^+=g(k(u))$ converges globally, and solutions of
the closed-loop system are bounded,
\ei
yet a stable limit-cycle oscillation exists in the closed-loop system.
This establishes, by means of a simple counterexample, that {\em monotonicity}
of the open-loop system is an essential assumption in our theorem.
Thus, robustness is only guaranteed with respect to uncertainty that preserves
monotonicity of the system.

The idea underlying the construction is very simple.  The open-loop system is
linear, and has the following transfer function:
\[
W(s) = \frac{-s+1}{s^2+(0.25)s+1}\,.
\]
Since the DC gain of this system is $W(0)=1$, and the system is stable, there
is a well-defined and increasing characteristic $k(u)=u$.  However a negative
feedback gain of $1/2$ destabilizes the system, even though the discrete
iteration $u^+=(-1/2)u$ is globally convergent.  (The $H_\infty $ gain of the
system is, of course, larger than $1$, and therefore the standard small-gain
theorem does not apply.)
In state-space terms, we use this system:
\beqn
\dot x_1 &=& (-1/4)x_1 - x_2 + 2u\\
\dot x_2 &=& x_1\\
    y &=& (1/2)(x_2-x_1) \,.
\eeqn
Note that, for each constant input $u\equiv u_0$, the solution of the system
converges to $(0,u_0/2)$, and therefore the output converges to $u_0$, so
indeed the characteristic $k$ is the identity.

We only need to modify the feedback law in order to make solutions of the
closed-loop globally bounded.  
For the feedback law we pick $g(x) = -0.5 \mbox{sat}(y)$, where
$\mbox{sat}(\cdot):=\mbox{sign}(\cdot)\min\{1,|\cdot|\}$ is a
saturation function.
The only equilibrium of the closed-loop system is at $(0,0)$.

The discrete iteration is
\[
u^+=-(1/2)\mbox{sat}(u)\,.
\]
With an arbitrary initial condition $u_0$, we have that
$u_1=-(1/2)\mbox{sat}(u_0)$, so that $|u_1|\leq 1/2$.
Thus $u_k=(-1/2)u_{k-1}$ for all $k\geq 2$, and indeed $u_k\rightarrow 0$ so
global convergence of the iteration holds.

\begin{figure}[ht]
\begin{center}
\includegraphics[scale=0.5]{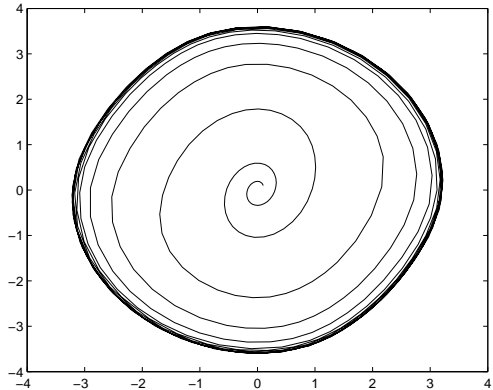}
\caption{Limit cycle in counterexample}
\label{fig:counter}
\end{center}
\end{figure}

However, global convergence to equilibrium fails for the closed-loop system,
and in fact there is a periodic solution.
Indeed, note that trajectories of the closed loop system are bounded, because
they can be viewed as solutions of a stable linear system forced by a bounded
input.  Moreover, since the equilibrium is a repelling point, it follows by the
Poincar\'e-Bendixson Theorem that a periodic orbit exists.
Figure~\ref{fig:counter} is a simulation showing a limit cycle.

\newcommand{\book}[1]{{\em #1\/},}
\newcommand{\inbook}[1]{in {\em #1\/},}
\newcommand{\journal}[1]{{\em #1\/}}
\newcommand{\Title}[1]{``#1,''}
\newcommand{\jvol}[1]{{\bf #1}}
\newcommand{\jyear}[1]{(#1),}
\newcommand{\pp}[1]{pp.\ #1}
\newcommand{\AND}{}
\newcommand{\auth}[2]{#1, #2,}

\end{document}